\documentclass[aoas,nameyear,rotating,seceqn,dvips]{arximspdf}
\usepackage{dcolumn}
\usepackage{graphicx}

\doi{10.1214/09-AOAS290}
\volume{4}
\issue{1}
\pubyear{2010}
\firstpage{460}
\lastpage{483}

\makeatletter
\renewcommand{\vec}[1]{\bolds{#1}}
\newcolumntype{d}[1]{D{.}{.}{#1}}
\newcolumntype{t}[1]{D{.}{ }{#1}}

\def\T{{\mathrm{\tiny T}}}

\makeatother

\begin{document}
\begin{frontmatter}

\title{Estimation of constant and time-varying dynamic parameters of
HIV infection in a nonlinear differential equation model\protect\thanksref{T1}}
\pdftitle{Estimation of constant and time-varying dynamic parameters of
HIV infection in a nonlinear differential equation model}
\thankstext{T1} {Supported in part by NIAID/NIH Grants AI055290,
AI50020, RR024160 and AI078498. Liang's research was
supported in part by the NSF Grant DMS-08-06097 and Miao's research
was supported in part by the Provost's Multidisciplinary Award and
the DCFAR Mentoring Award from the University of Rochester.}
\runtitle{Parameter Estimation in Nonlinear ODE Models}

\begin{aug}
\author[a]{\fnms{Hua} \snm{Liang}\ead[label=e1]{hliang@bst.rochester.edu}\corref{}},
\author[a]{\fnms{Hongyu} \snm{Miao}\ead[label=e2]{Hongyu\_Miao@urmc.rochester.edu}}
\and
\author[a]{\fnms{Hulin} \snm{Wu}\ead[label=e3]{hwu@bst.rochester.edu}}
\runauthor{H. Liang, H. Miao and H. Wu}
\affiliation{University of Rochester}
\address[a]{Department of Biostatistics\\
\quad and Computational Biology\\
University of Rochester\\
School of Medicine\\
\quad and Dentistry\\
601 Elmwood Avenue, Box 630\\
Rochester, New York 14642 \\USA\\
\printead{e1}}
\end{aug}

\received{\smonth{4} \syear{2009}}
\revised{\smonth{9} \syear{2009}}

%
\begin{abstract}

Modeling viral dynamics in HIV/AIDS studies has resulted in a deep
understanding of pathogenesis of HIV infection from which novel
antiviral treatment guidance and strategies have been derived. Viral
dynamics models based on nonlinear differential equations have been
proposed and well developed over the past few decades. However, it is
quite challenging to use experimental or clinical data to estimate the
unknown parameters (both constant and time-varying parameters) in
complex nonlinear differential equation models. Therefore,
investigators usually fix some parameter values, from the literature or
by experience, to obtain only parameter estimates of interest from
clinical or experimental data. However, when such prior
information is not available, it is desirable to determine all the
parameter estimates from data. In this paper we intend to combine the
newly developed approaches, a multi-stage smoothing-based (MSSB) method
and the spline-enhanced nonlinear least squares (SNLS) approach, to
estimate all HIV viral dynamic parameters in a nonlinear differential
equation model. In particular, to the best of our knowledge, this is
the first attempt to propose a comparatively thorough procedure,
accounting for both efficiency and accuracy, to rigorously
estimate all key kinetic parameters in a nonlinear differential
equation model of HIV dynamics from clinical data. These parameters
include the proliferation rate and death rate of uninfected
HIV-targeted cells, the average number of virions produced by an
infected cell, and the infection rate which is related to the antiviral
treatment effect and is time-varying. To validate the estimation
methods, we verified the identifiability of the HIV viral dynamic model
and performed simulation studies. We applied the proposed techniques to
estimate the key HIV viral dynamic parameters for two individual AIDS
patients treated with antiretroviral therapies. We demonstrate that HIV
viral dynamics can be well characterized and quantified for individual
patients. As a result, personalized treatment decision based on viral
dynamic models is possible.

\end{abstract}

%

\end{frontmatter}
%

\section{Introduction}\label{sec:intr}

The AIDS pandemic continues to pose a serious threat to public health
worldwide. AIDS has killed an estimated 2.9 million people, with close
to 4.3 million people newly infected with HIV in 2006 alone. The total
number of people living with HIV has reached its highest level (WHO web
site). Although highly active antiretroviral therapy (HAART) regimens
are effective in suppressing plasma HIV RNA levels (viral load) below
the limit of detection, many patients may fail HAART treatments due to
drug resistance, lack of potency, poor drug adherence, pharmacokinetic
problems and adverse effects. In addition, the complexity of regimens
and lack of full understanding of the pathogenesis of HIV infection
also pose great challenges to AIDS researchers. Over the past 2
decades, many mathematicians and statisticians have developed
mechanism-based models and statistical approaches to assist in
understanding HIV pathogenesis and have made significant contributions
in this area [Ho et al. (\citeyear{Ho1995}), Wei et al. (\citeyear
{Wei1995}), Perelson et al.
(\citeyear{Perelson1996,Perelson1997}), Wu et al. (\citeyear
{Wuetal1999})]. In particular,
differential equation models have been widely used in describing
dynamics and interactions of HIV and the immune system. Some survey on
these models can be found in Perelson and Nelson (\citeyear
{Perelson1999}), Nowak and May
(\citeyear{Nowak2000}), and Tan and Wu (\citeyear{Tan2005}).

For the mechanism-based models of HIV infection, one critical question
is how to use experimental or clinical data to estimate the parameters
in the nonlinear differential equation models which do not have
closed-form solutions in most cases. Researchers have made a
substantial effort to get an approximate closed-form solution under
various assumptions, and then use the standard regression approach to
estimate the dynamic parameters in the models [Ho et al. (\citeyear
{Ho1995}), Wei et
al. (\citeyear{Wei1995}), Perelson et al.
(\citeyear{Perelson1996,Perelson1997}), Wu et al. (\citeyear
{Wuetal1999}), Wu and Ding
(\citeyear{WuDing1999}),
Wu (\citeyear{Wu2005})]. But these approximations and assumptions may
not always
hold, in particular, for patients undergoing long-term treatment.

In this paper we consider the following HIV dynamic model for patients
under long term treatments [e.g., Perelson and Nelson (\citeyear
{Perelson1999})]:
\begin{eqnarray} \label{3Dmodel}
\frac{d}{dt}T_U(t) &=& \lambda- \rho T_U(t) - \eta(t) T_U(t) V(t),
\label{3Dmodel:Eq1} \\
\frac{d}{dt}T_I(t) &=& \eta(t) T_U (t) V(t) - \delta T_I(t), \label
{3Dmodel:Eq2} \\
\frac{d}{dt}V(t) &=& N \delta T_I(t)- c V(t), \label{3Dmodel:Eq3}
\end{eqnarray}
where $T_U$ is the concentration of uninfected target \mbox{CD4$+$ T} cells,
$T_I$ the concentration of infected cells, $V$ the viral load, $\lambda
$ the proliferation rate of uninfected target cells, $\rho$ the death
rate of uninfected target cells, $\eta$(t) the time-varying infection
rate depending on antiviral drug efficacy, $\delta$ the death rate of
infected cells, $c$~the clearance rate of free virions, and $N$ the
number of virions produced by a single infected cell on average.
In this system, $T_U(t)$, $T_I(t)$ and $V(t)$ are state variables, and
$(\lambda,\rho,N,\delta,c,\eta(t))^{\T}$ are unknown dynamic
parameters. Notice that here we do not distinguish the infected cells
by various subpopulations such as productively infected cells, latently
infected cells and long-lived infected cells [Perelson et al.
(\citeyear{Perelson1996,Perelson1997})],
since we intend to model the viral dynamics for an HIV-infected
patient under long-term treatment. We also use a time-varying parameter
$\eta(t)$ to model the infection rate since the infection rate may
change nonparametrically due to the variation in treatment effect over
time. Generally speaking, all kinetic parameters in this model could be
time-varying [not only $\eta(t)$]; however, it is usually unfeasible to
do so due to the limited data collected in the clinical studies. Thus,
this model provides a flexible although simple approach for studying
long-term viral dynamics. In clinical trials or clinical practice,
viral load, $V(t)$ and total \mbox{CD4$+$ T} cell count, $T(t)=T_U(t)+T_I(t)$,
are closely monitored and measured over time.

In practice, some parameters can be fixed from previous studies and
only the remaining parameters are needed to be estimated. However, when
such prior knowledge is not available, it is important to estimate all
viral dynamic parameters, $(\lambda,\rho,N,\delta,c,\eta(t))^{\T
}$, for
each individual patient from the clinical measurements of $V(t)$ and
$T(t)$, since the estimated dynamic parameters may be used to guide
clinical decisions for individualized treatment. Although AIDS
investigators have tried to estimate some of these dynamic parameters
based on viral load data
[Ho et al. (\citeyear{Ho1995}), Wei et al. (\citeyear{Wei1995}),
Perelson et al.
(\citeyear{Perelson1996,Perelson1997}),
Wu et al. (\citeyear{Wuetal1999})], none of them have successfully
estimated all these dynamic parameters directly for an individual
patient. Huang and Wu (\citeyear{HuangWu2006}) and Huang, Liu and Wu
(\citeyear{Huangetal2006}) have made an
effort to use the Bayesian method to estimate all these parameters, but
strong priors are required for most parameters in order to make all
parameters identifiable. In the work of Xia (\citeyear{Xia2003}),
Filter, Xia and
Gray (\citeyear{Filter2005}), Gray et al. (\citeyear{Gray2005}), and
Ouattara, Mhawej and Moog (\citeyear{Ouattara2008}),
all model parameters were estimated for HIV dynamic models with only
constant coefficients. In this paper, to the best of our knowledge,
this effort represents the first attempt to simultaneously estimate all
the constant and time-varying HIV viral dynamic parameters for
individual patients using both viral load and total \mbox{CD4$+$ T} cell count data.

In this paper we propose two estimation approaches. The first one is a
multistage smoothing-based (MSSB) approach. We derive the direct
relationships between the unknown dynamic parameters and measurement
variables from the original differential equation models, and then we
employ these relationships to formulate regression models for the
unknown parameters. These regression models involve not only the time
function of measurement variables, but also their derivatives. We
propose using a nonparametric smoothing method, such as the local
polynomial smoothing, to obtain the measurement variables and their
derivatives which are substituted into the regression models to
estimate the unknown dynamic parameters (including the time-varying
parameter) in the second step. This approach avoids directly solving
the ODEs by numerical methods and is computationally efficient. The
second approach that we propose is to use a spline-enhanced nonlinear
least squares (SNLS) method. This approach is to use splines to
approximate the time-varying parameter so that the original ODE model
becomes a model only containing constant parameters. The standard NLS
method can be employed to estimate the unknown dynamic parameters and
spline coefficients. This approach needs
to use a numerical method to repeatedly solve the nonlinear ODEs for
the high-dimensional NLS optimization problem which is computationally
challenging. Some cutting-edge computational techniques are necessary
to solve this problem. Note that the MSSB approach is computationally
efficient, but the derived estimates are quite rough; however, these
estimates can be used as the initial values for the SNLS method so that
the two approaches can be efficiently combined for practical applications.

The rest of this paper is organized as follows. In Section \ref
{multi-stage} we briefly
discuss the identifiability of the HIV dynamic model, then propose the
multistage smoothing-based (MSSB) approach. In Section \ref{sec3} we
introduce
the spline-enhanced nonlinear least squares (SNLS) method. In
particular, a hybrid optimization technique combining a gradient method
and a global optimization algorithm will be used to tackle the
high-dimensional NLS optimization problem. The simulation studies are
presented to evaluate the performance of the proposed approaches in
Section \ref{sec4}. We apply the proposed methods to estimate HIV
viral dynamic
parameters (including the time-varying infection rate) from data for
two AIDS patients in Section \ref{SecClinicalData}. We demonstrate
that some of these
dynamic parameters are estimated from clinical data for the first time.
We conclude the paper with some discussions in Section \ref{sec6}.
The details
of the proposed hybrid optimization algorithm are given in the
\hyperref[app]{Appendix}.

\section{A multistage smoothing-based (MSSB) approach} \label{multi-stage}

Before introducing the estimation methods, it is critical to determine
whether all unknown model parameters can be uniquely and reliably
identified from a given system input and measurable output. The
technique to answer this question is called identifiability analysis.
There are mainly two types of identifiability problems: structural
(mathematical) identifiability and practical (statistical)
identifiability. Structural identifiability analysis techniques use
model structure information only (without knowledge of experimental
observations) to determine whether all unknown
parameters are uniquely identifiable. Therefore, it is also called the
prior identifiability analysis. In contrast, practical identifiability
analysis techniques should be applied after model fitting to verify the
reliability of estimates, so it is called the posterior identifiability
analysis [Rodriguez-Fernandez, Egea and Banga (\citeyear
{Rodriguez2006})]. A~thorough
discussion and review of identifiability techniques and theories is out
of the scope of this study; for more details, the interested reader is
referred to Ritt~(\citeyear{Ritt1950}), Bellman and {\AA}str\"{o}m
(\citeyear{Bellman1970}), Kolchin
(\citeyear{Kolchin1973}), Pohjanpalo (\citeyear{Pohjanpalo1978}),
Cobelli, Lepschy and Jacur (\citeyear{Cobelli1979}), Walter (\citeyear
{Walter1987}),
Vajda, Godfrey and Rabitz (\citeyear{Vajda1989}), Ollivier (\citeyear
{ollivier1990}), Chappel and Godfrey (\citeyear{Chappel1992}), Ljung and
Glad (\citeyear{Ljung1994}), Audoly et al. (\citeyear{Audoly2001}),
Xia and Moog (\citeyear{Xia2003}), Jeffrey and Xia
(\citeyear{Jeffrey2005}), Wu et al. (\citeyear{Wu2008}) and Miao et
al. (\citeyear{Miao2008}). In this study we
employed the differential algebra approach [Ljung and Glad (\citeyear
{Ljung1994})] and
it can be easily verified that all constant and time-varying parameters
in models~(\ref{3Dmodel:Eq1})--(\ref{3Dmodel:Eq3}) are structurally
identifiable.

Parameter estimation for ODE models has been investigated using the
least squares principle by mathematicians [Hemker (\citeyear
{Hemker1972}), Bard
(\citeyear{Bard1974}),
Li, Osborne and Prvan (\citeyear{Li2005})], computer scientists [Varah
(\citeyear{Varah1982})] and
chemical engineers [Ogunnaike and Ray (\citeyear{Ogunnaike1994}),
Poyton et al. (\citeyear{Poyton2006})].
Mathematicians have focused on the development of efficient and stable
algorithms to solve the least squares problem. Recently statisticians
have started to develop various statistical methods to estimate dynamic
parameters in ODE models. For example, Putter et al. (\citeyear
{Putter2002}), Huang and
Wu (\citeyear{HuangWu2006}), and Huang, Liu and
Wu (\citeyear{Huangetal2006}) have developed hierarchical Bayesian
approaches to estimate
dynamic parameters in HIV dynamic models for longitudinal data. Li et
al. (\citeyear{Li2002}) proposed a spline-based approach to estimate
time-varying
parameters in ODE models. Ramsay (\citeyear{Ramsay1996}) proposed a
technique named
principal differential analysis (PDA) for estimation of differential
equation models [see a comprehensive survey in Ramsay and Silverman
(\citeyear{Ramsay2005})]. Recently Ramsay et al. (\citeyear
{Ramsay2007}) applied a penalized spline
method to estimate the constant dynamic parameters in ODE models. Chen
and Wu (\citeyear{Chen2008,Chen2009}) and Liang and Wu (\citeyear
{Liang2008}) proposed a two-step
smoothing-based approach to estimate both constant and
time-varying parameters in ODE models separately.

In this section we adopt the ideas from Liang and Wu (\citeyear{Liang2008})
and Chen and Wu (\citeyear{Chen2008,Chen2009}) to use
the smoothing-based approach to estimate all dynamic parameters,
including both constant and time-varying parameters, in model~(\ref
{3Dmodel:Eq1})--(\ref{3Dmodel:Eq3}) in three stages. Stage I is to
smooth the noisy data to estimate the state variables and their
derivatives; Stage II is to estimate constant dynamic parameters
($\lambda, \rho, c$); Stage III is proposed to estimate both constant
dynamic parameters, $\delta$ and $N$, and the time-varying parameter
$\eta(t)$. We introduce these three stages in detail in the following
subsections.

\subsection{Stage I: Local polynomial estimation of the state
variables} \label{Sect2.1}

In this subsection we briefly describe how we use the local polynomial
regression technique to estimate the functions $T(t)$ and $V(t)$ and
their derivatives. Consider a general situation that a state variable,
$X(t)$, is observed at $n$ time points, $(Y_1, \ldots, Y_n)$, that is,
$Y_i=X(t_i)+e_i$ for $i=1,\ldots, n$, where $(e_1, e_2, \ldots, e_n)$
are independent measurement errors with mean zero.
Assume that the fourth derivative of $ X(t)$ exists. For each given
time point~$t_0$, we approximate the function $X(t_i)$\vspace*{1pt} locally by a $p${th}-order
polynomial,
$X(t_i)\approx X(t_{0})+(t_i-t_0)X^{(1)}(t_0)+\cdots
+X^{(p)}(t_0)(t_i-t_0)^p/{p!}
\triangleq\sum_{j=0}^p\zeta_j(t_0)(t_i-t_0)^j$,
for $t_i, i = 1,\ldots,n,$ in a neighborhood of the point~$t_0$, where
$\zeta_j(t_{0})= X^{(j)}(t_{0})$ for $j= 0,1,\ldots, p$. Following the
local polynomial fitting [Fan and Gijbels (\citeyear{Fan1996})], the
estimators $\widehat
X^{(\nu)}(t)$ of $X^{(\nu)}(t)$ ($\nu=0, 1, 2$ in our case) can be
obtained by minimizing the locally weighted least-squares criterion,
\[
\sum_{i=1}^n\Biggl\{Y_i-\sum_{j=0}^p\zeta_j(t_i-t)^j\Biggr\}^2K_h(t_i-t),
\]
where $K(\cdot)$ is a symmetric kernel function, $K_h(\cdot)=K(\cdot
/h)/h$, and $h$ is a proper bandwidth. Let $ Y=(Y_1,\ldots,Y_n)^\T$
denote $\mathbf{T}_{p,t}$ as an $n\times(p+1)$ design matrix whose
$i${th} row is $(1, t_i-t, \ldots, (t_i-t)^p)^\T$, and $\mathbf
{W}_t$ an
$n\times n$ diagonal matrix of kernel weights, that is, $\operatorname
{diag}\{
K_h(t_1-t),\ldots,K_h(t_n-t)\}$. Assuming that the matrix $\mathbf
{T}_{p,t}^\T\mathbf{W}_t\mathbf{T}_{p,t}$ is not singular, a standard
weighted least
squares theory leads to the solution, $\widehat{\bolds{\zeta}}=(\mathbf{T}_{p,t}^\T\mathbf{W}
_t\mathbf{T}_{p,t})^{-1}\mathbf{T}_{p,t}^\T\mathbf{W}_t Y,$ where
$p=1, 2, 3$. We use the local linear regression to estimate $X(t)$, the
local quadratic regression to estimate $X'(t)$, and the local cubic
regression to estimate $X^{(2)}(t)$. Consequently, using the above
notation, the estimators $\widehat X^{(q)}(t)$ can be expressed as
\[
\widehat {X}^{(q)}(t)=\bolds{\xi}_{(q+1)}^\T(\mathbf{T}_{1+q,t}^\T
\mathbf{W}_t\mathbf{T}_{1+q,t})^{-1}
\mathbf{T}_{1+q,t}^\T\mathbf{W}_t Y\qquad \mbox{for } q=0,1,2,
\]
where $\bolds{\xi}_q$ is the $(q+2)\times1$ vector having 1 in the $(q+1)$
entry and zeros in the other entries.

We apply the local polynomial smoothing technique to the viral load
data, $V(t)$ and total CD4$+$ T cell count data, $T(t)=T_U(t)+T_I(t)$ to
obtain the estimates of $\widehat V(t)$, $\widehat T(t)$, $\widehat
V'(t)$ and $\widehat
T'(t)$ that will be used in Stage II, and the estimate of $V^{(2)}(t)$
that will also be used in Stage III.

\subsection{Stage II: PsLS estimates}

In this subsection we apply the approach proposed by Liang and Wu
(\citeyear{Liang2008}) to estimate constant dynamic parameters
($\lambda, \rho, c$) in
models~(\ref{3Dmodel:Eq1})--(\ref{3Dmodel:Eq3}). Notice that we can
combine equations (\ref{3Dmodel:Eq1}) and (\ref{3Dmodel:Eq2}), and obtain
\[
\frac{d}{dt}[T_U(t)+T_I(t)]=\lambda- \rho T_U(t)- \delta T_I(t).
\]
Recalling that $T(t)=T_I(t) + T_U(t)$ and substituting
$T_U(t)=T(t)-T_I(t)$ in the above equation, we obtain
\[
\frac{d}{dt}T(t)=\lambda- \rho[T(t)-T_I(t)]- \delta T_I(t).
\]
From the above equation, we obtain
\[
T_I=\frac{-\lambda}{\rho-\delta}+\frac{\rho}{\rho-\delta
}T+\frac{1}{\rho
-\delta}T',
\]
where $T'=dT(t)/dt$.\vspace*{1pt} Substituting this expression into equation (\ref
{3Dmodel:Eq3}) and letting $\alpha_0=-\frac{N\delta\lambda}{\rho
-\delta}$, $\alpha_1=\frac{N\delta\rho}{\rho-\delta}$ and
$\alpha
_2=\frac{N\delta}{\rho-\delta}$, we have
\begin{equation}
V'(t)=\alpha_0+\alpha_1 T(t)+\alpha_2 T'(t)-c V(t), \label{Eq4}
\end{equation}
where $V(t)$ and $T(t)$ for $t=t_1, t_2, \ldots, t_n$ are the
measurements of viral load and \mbox{CD4$+$ T} cell count from clinical studies
which are subject to measurement errors.

Let $\widehat V(t)$, $\widehat T(t)$, $\widehat V'(t)$ and
$\widehat T'(t)$ be the
estimates of $V(t)$, $T(t)$, $V'(t)$ and $T'(t)$ from Section~\ref
{Sect2.1}, respectively, and substitute these estimates in (\ref
{Eq4}). We then obtain a linear regression model
[Liang and Wu (\citeyear{Liang2008})],
\begin{equation}
\widehat V'(t)=\alpha_0+\alpha_1 \widehat T(t)+\alpha_2 \widehat
T'(t)-c \widehat
V(t)+\varepsilon (t), \label{eq:Eq4-1}
\end{equation}
where $\varepsilon (t)$ includes all substitution errors. Using the least
squares regression technique, we obtain the estimates of $\alpha_0$,
$\alpha_1$ and $\alpha_2$ which are termed as the pseudo-least squares
(PsLS) estimates in Liang and Wu~(\citeyear{Liang2008}). Notice that,
from the above
derivation, we have the relationship $\lambda=-\alpha_0/\alpha_2$ and
$\rho=\alpha_1/\alpha_2$. Thus, we can obtain the estimates of $c$,
$\lambda$ and $\rho$ from model (\ref{eq:Eq4-1}).

Write $\bolds{\alpha}=(\alpha_0, \alpha_1, \alpha_2, c)$ and $
\bolds{\Lambda}
=\operatorname{diag}(1,\widehat{T}, \widehat{T}', -\widehat{V})$.
Let $\mu_l(K)=\break\int
_{-1}^1 z^l K(z)\,dz$ for $l=0,1,2.$ Using the arguments similar to
those in
Liang and Wu~(\citeyear{Liang2008}),
we can establish asymptotics for the smoothing-based estimators
$\widehat{\bolds{\alpha}}$ of~$\bolds{\alpha}$.
%
The Delta-method can then be used for establishment of the asymptotic
properties of the estimators for $\lambda$ and $\rho$. Notice that,
when the mean of $\varepsilon (t)$ is zero, the least squares
estimates of
$\alpha_0$, $\alpha_1$ and $\alpha_2$ are unbiased. However, the
substitution error $\varepsilon (t)$ in model (\ref{eq:Eq4-1}) is not
mean zero,
instead its mean is in the order of $h^2$, and variance is
of the order $(nh)^{-1}$ which goes to zero when $n \rightarrow\infty
$. Thus, $\varepsilon (t)$ is different from a standard measurement
error with mean zero and constant variance.

\subsection{Stage III: Semiparametric regression for estimation of both
constant and time-varying parameters}

Now we propose how to estimate the remaining constant parameters, $N$
and $\delta$, and the time-varying parameter, $\eta(t)$. Recalling
(\ref{3Dmodel:Eq2}), we have
\begin{equation}
\frac{d}{dt}T_I(t)=\eta(t)\{T(t)-T_I(t)\}V(t)-\delta T_I(t), \label
{eq:7}
\end{equation}
while from (\ref{3Dmodel:Eq3}) we can obtain
\begin{equation}
\frac{d}{dt}T_I(t)=\frac{V^{(2)}(t)+cV'(t)}{N\delta}. \label{eq:8}
\end{equation}
A combination of (\ref{eq:7}) and (\ref{eq:8}) deduces
\begin{eqnarray}\label{eq:9}
V^{(2)}(t)+cV'(t)&=&\eta(t) N\delta T(t)
V(t)-\eta(t)\{V'(t)V(t)+cV^2(t)\}
\nonumber
\\[-8pt]
\\[-8pt]
\nonumber
&&{}-\delta\{V'(t)+cV(t)\}.
\end{eqnarray}
Let\vspace*{1pt} $Z(t)=\widehat V^{(2)}(t)+\widehat c\widehat V'(t)$, $U_1(t)=-\{
\widehat V'(t)\widehat
V(t)+\widehat c\widehat V^2(t)\}$,
$U_2(t)=-\{\widehat V'(t)+\widehat c\widehat V(t)\}$,
$U_3(t)=\widehat T(t)\widehat V(t),$ in
which all functions and parameters have been estimated from Stages I
and II. Thus, from (\ref{eq:9}), we can formulate a semiparametric
time-varying coefficient regression model [Wu and Zhang (\citeyear{Wu2006})]:
\begin{equation}\label{eq:9-1}
Z(t)=U_1(t)\delta+U_2(t) \eta(t)+U_3(t) N\delta\eta(t)+
\varepsilon ^*(t),
\end{equation}
where
$\varepsilon ^*(t)$ includes all substitution errors. Note that
$\delta$ and
$N\delta$ are constant parameters, while $\eta(t)$ is an unknown
time-varying parameter. To fit this model, we can approximate the
time-varying parameter $\eta(t)$ by the local polynomial approach,
a~basis spline method or other nonparametric techniques [Wu and Zhang
(\citeyear{Wu2006})]. The basis spline approach is straightforward and simple if the
function $\eta(t)$ does not fluctuate dramatically as in our case of
HIV dynamic models. Thus, we select to use the B-spline approximation
here, that is,
\begin{equation}
\label{etabasisspline}
\eta(t) \approx\sum_{j=1}^{s} a_j b_{j,k} (t),
\end{equation}
where $a_j$ are constant B-spline coefficients, and $b_{j,k}(t)$ the
basis functions of order $k$.
For more details about B-splines such as construction of higher order
basis functions via recurrence
relations, the reader is referred to de Boor (\citeyear{deBoor1978}). Thus, model (\ref
{eq:9-1}) can be approximately
written as
\begin{eqnarray}\label{eq:9-1-1}
 Z(t)&=&U_1(t)\delta+ \sum_{j=1}^{s} \{b_{j,k} (t) U_2(t)
\} a_j
\nonumber
\\[-8pt]
\\[-8pt]
\nonumber
&&{}+\sum_{j=1}^{s} \{b_{j,k} (t) U_3(t)\}
N\delta a_j+ \varepsilon ^*(t).
\end{eqnarray}
This becomes a standard linear regression model and we can easily
estimate the unknown constant parameters $\delta, a_j$ and $N\delta
a_j$ $(j=1,2, \ldots, s)$ from which we can derive the estimates of $\delta,
N$ and $\eta(t)$. Note that AIC, BIC or AICc can be used to determine
the order of the B-splines in our numerical data analysis. See detailed
discussion on this in the next section.

Also notice that Wu and Ding (\citeyear{WuDing1999}) suggested that fitting viral
dynamic data to different models for different time periods may result
in better estimates of viral dynamic parameters. For example, we may
fit a viral dynamic model with a closed-form solution to the early
(first week) viral load data to obtain a better estimate of parameters
$\delta$ and $c$ as proposed by Perelson et al. (\citeyear{Perelson1996}). These estimates
can then be substituted into the regression models in Stages II and III
to obtain the estimates of other parameters. We will adopt this
alternative strategy in our real data analysis.

In summary, we are able to use a multistage approach to derive the
estimates of all dynamic parameters in an HIV dynamic model. This
approach only involves parametric and nonparametric/semiparametric
regressions and the implementation is straightforward. The numerical
evaluations of ODEs are avoided and the initial values of the state
variables are not required. However, there are two issues that should
be addressed for this approach. First, in this study, we focused on
models~(\ref{3Dmodel:Eq1})--(\ref{3Dmodel:Eq3}) and the three stages
were therefore particularly proposed for the specific model. However,
the MSSB approach itself is a general method for estimating parameters
in ODE models; for general guidelines in use of the MSSB approach, the
interested reader is referred to Liang and Wu (\citeyear{Liang2008}). Second, although
the estimators based on the MSSB approach have some attractive
asymptotic properties, some limitations exist for this approach. In
particular, it needs to estimate the derivatives of state variables
which are sensitive to measurement noise when the data are sparse. This
may result in biased estimates of parameters and measurement errors. In
order to improve the estimates, we propose the spline-enhanced least
squares approach in the next section.

\section{The spline-enhanced nonlinear least squares (SNLS)
approach}\label{sec3}

In this section we introduce a spline-enhanced nonlinear least squares
(SNLS) method to refine parameter estimates for models~(\ref
{3Dmodel:Eq1})--(\ref{3Dmodel:Eq3}). The basic idea is to approximate
the time-varying parameter $\eta(t)$ by the B-spline approach. Then
models~(\ref{3Dmodel:Eq1})--(\ref{3Dmodel:Eq3}) become the standard
ODEs with constant parameters only. We can use the standard NLS
approach to estimate the constant dynamic parameters and B-spline
coefficients by numerically evaluating the ODEs repeatedly. This method
is computationally intensive. Li et al. (\citeyear{Li2002}) used a similar idea to
approximate the unknown time-varying parameter in a pharmacokinetic ODE model.

\subsection{Spline approximation and nonlinear least squares}
\label{splinemodelselect}

Different types of splines can be constructed based on different basis
functions, such as the well-known piecewise polynomial splines and
basis splines. Note that, for an arbitrary spline function of a
specific degree and smoothness over a given domain partition, a linear
combination of basis splines of the same degree and smoothness over the
same partition can always be found to represent this spline function
[de Boor~(\citeyear{deBoor1978})]. Therefore, B-splines can be employed to approximate
the time-varying parameter $\eta(t)$ in this study without loss of generality.

Similar to the approximation (\ref{etabasisspline}) in the previous
section, $\eta(t)$ can be approximated by a B-spline of order $k$ with
$s$ control points [de Boor (\citeyear{deBoor1978})], that is, $\eta(t) \approx\sum
_{j=1}^{s} a_j b_{j,k} (t).$ In addition, it should be noted that once
the number and positions of control points are determined, the number
and positions of knots are also automatically determined at the knot
average sites [de Boor (\citeyear{deBoor1978})]. Thus, our model equations~(\ref
{3Dmodel:Eq1})--(\ref{3Dmodel:Eq3}) can be approximated by
\begin{eqnarray}
\frac{d}{dt}T_U(t) &=& \lambda- \rho T_U(t) - \Biggl\{\sum_{j=1}^{s} a_j
b_{j,k}(t)\Biggr\} T_U(t)
V(t), \label{3DmodelEtaSpline:Eq1} \\
\frac{d}{dt}T_I(t) &=& \Biggl\{\sum_{j=1}^{s} a_j b_{j,k}(t)\Biggr\} T_U
(t) V(t) - \delta T_I(t),
\label{3DmodelEtaSpline:Eq2} \\
\frac{d}{dt}V(t) &=& N \delta T_I(t)- c V(t), \label{3DmodelEtaSpline:Eq3}
\end{eqnarray}
where $\vec{\theta}=(\lambda,\rho,N,\delta,c,a_1,a_2,\ldots
,a_s)^{\T}$
are unknown constant parameters. Note that we have measurements of
total \mbox{CD4$+$ T} cell counts, $T(t_i)=T_U(t_i)+T_I(t_i)$, and viral load,
$V(t_i)$, that is, the measurement model can be written as
\begin{eqnarray}
\label{log10MeasurementModel}
Y_{1i} &=& T(t_i) + \varepsilon_{1i},\qquad i=1,2,\ldots,n_T,\\
Y_{2j} &=& V(t_j) + \varepsilon_{2j}, \qquad j=1,2,\ldots,n_V, \label
{log10MeasurementModel4.3}
\end{eqnarray}
where $\varepsilon_{1i}$ and $\varepsilon_{2j}$ are assumed to be
mean zero
with constant variances and independent. Then the standard NLS
estimator can be obtained by minimizing the objective function
\begin{equation}
\label{log10RSS} \mathit{RSS}(\vec{\theta}) = \sum_{i=1}^{n_T} \{Y_{1i}
-T(t_i, \vec{\theta})\}^2 +
\sum_{j=1}^{n_V} \{Y_{2j} - V(t_j,\vec{\theta})\}^2,
\end{equation}
where $n_T$ is the total number of \mbox{CD4$+$ T} cell measurements and $n_V$
is the total number of viral load observations; and $T(t_i, \vec
{\theta
})$ and $V(t_j,\vec{\theta})$ are numerical solutions to
equations~(\ref{3DmodelEtaSpline:Eq1})--(\ref{3DmodelEtaSpline:Eq3})
using the Runge--Kutta method. However, if $\varepsilon_{1i}$ and~$\varepsilon
_{2j}$ are correlated, the weighted (generalized) NLS method can be
used. To stabilize the variance and the computational algorithms, the
log-transformation of the data is usually used in practice. More
generally, the weighted NLS approach can be used to more efficiently
estimate the unknown parameters if the weights for different terms in
the objective function are known. Usually the Fisher-Information-Matrix
can be used to obtain the confidence intervals for the unknown
parameters, but the bootstrap approach is more precise although it is
more computationally intensive [Joshi, Seidel-Morgenstern and Kremling
(\citeyear{Joshi2006})]. We will use the bootstrap method in our real data analysis.

\subsection{Hybrid optimization and spline parameter selection}

For the SNLS approach, a critical step is to minimize a multimodal
objective function (\ref{log10RSS}) over a~high-dimensional parameter
space. In practice, it is challenging to find the global solution to
such problems if the parameter space is high-dimensional (as always the
case in many biomedical problems), and/or the parameter values are of
different orders of magnitude, and/or the objective function is
multimodal or even not smooth with noisy data. Thus, it is critical to
develop an efficient and stable optimization algorithm.

There are three main categories of optimization methods: direct search
methods, gradient methods and global optimization methods. However,
both the direct search methods (e.g., the Simplex method) and the
gradient methods (e.g, the Levenberg--Marquardt method and the
Gauss--Newton method) can be easily trapped by local minima and even
just fail for our problems [Miao et al. (\citeyear{Miao2008})]. For details and
application of such methods in ODE models, the reader is referred to
Nocedal and Wright (\citeyear{Nocedal1999}) and Englezos and Kalogerakis (\citeyear{Englezos2001}).
Therefore, the global optimization methods are more suitable for the
parameter estimation problem for ODE models. Moles, Banga and Keller
(\citeyear{Moles2004}) compared the performance and computational cost of seven global
optimization methods, including the differential evolution method
[Storn and Price (\citeyear{Storn1997})]. Their results suggest that the differential
evolution method outperforms the other six methods with a reasonable
computational cost. Unfortunately, existing global optimization methods
are very computationally intensive. Improved performance can be
achieved by combining gradient methods and global optimization methods,
called hybrid methods. A hybrid method based on the scatter search and
sequential quadratic programming (SQP) has been proposed by
Rodriguez-Fernandez, Egea and Banga~(\citeyear{Rodriguez2006}), who showed that the hybrid scatter
search method is much faster than the differential evolution method for
a simple HIV ODE model. However, our preliminary work did not show
improved performance of the scatter search method without SQP with
respect to the differential evolution method in terms of computational
cost and convergence rate. Thus, the performance improvement of the
hybrid scatter search method is mainly due to the incorporation of the
SQP local optimization method. In addition, Vrugt and Robinson (\citeyear{Vrugt2007}) suggested
that a significant improvement of efficiency can be achieved by
combining multiple global optimization algorithms. Here we combined the
differential evolution and the scatter search method and incorporated
with the SQP local optimization method [e.g., SOLNP by Ye (\citeyear{Ye1987})] for
parameter estimation of ODE models. We present the details of the
proposed algorithm in the \hyperref[app]{Appendix}.

Another critical problem that needs to be addressed in order to fit
models~(\ref{3DmodelEtaSpline:Eq1})--(\ref{3DmodelEtaSpline:Eq3}) is
to determine the order $k$ and the number and positions of control
points $s$ for the spline approximation. In general, we assume that
$\eta(t)$ is a first-order continuous function of time. Therefore,
B-splines of order 2 (piecewise straight lines) should not be
considered. Also, since the high order B-spline approximation (e.g., $k
\geq5$) may introduce unnecessary violent local oscillations (called
Runge's phenomenon) [Runge (\citeyear{Runge1901})], we consider up to 4{th} order
B-splines in this study. As to the positions of the control points, to
account for highly-skewed data, we select the control points' positions
such that they are equally-spaced in the logarithm time scale. We can
use AIC, BIC and AICc criteria [Akaike (\citeyear{Akaike1973}), Schwarz (\citeyear{Schwarz1978}), Burnham
and Anderson (\citeyear{Burnham2004})] to determine the order and the knots, that is,
\begin{eqnarray}
\label{generalAICBICAICc}
\mathit{AIC} &=& -2 \ln L + 2K, \\
\mathit{BIC} &=& -2 \ln L + K \ln(N), \\
\mathit{AICc} &=& \mathit{AIC} + \frac{2K(K+1)}{N-K-1},
\end{eqnarray}
where $L$ denotes the likelihood function, $N$ the total number of
observations and~$K$ the number of unknown parameters. Under the
normality assumption of measurement errors, these model selection
criteria can be rewritten as
\begin{eqnarray}
\label{LSAICBICAICc}
\mathit{AIC} &=& N \ln\biggl(\frac{\mathit{RSS}}{N}\biggr) + 2K, \\
\mathit{BIC} &=& N \ln\biggl(\frac{\mathit{RSS}}{N}\biggr) + K \ln(N), \\
\mathit{AICc} &=& N \ln\biggl(\frac{\mathit{RSS}}{N}\biggr) + \frac{2NK}{N-K-1},
\end{eqnarray}
where $\mathit{RSS}$ is the residual of sum squares obtained from the NLS model
fitting. It should also be noticed that, as the number of spline knots
goes to large, the associated parameter space will be high-dimensional.
In this case, the novel and efficient optimization methods such as the
hybrid optimization algorithms
are necessary to locate the global minima of the NLS objective
function. In addition, it is necessary to predetermine a good and
informative search range for each of the unknown parameters in order to
ease the computational burden of the proposed optimization algorithm.
Thus, it is necessary and a good strategy to combine the proposed two
approaches to estimate the dynamic parameters in a complex dynamic
system. That is, first the MSSB approach introduced in the previous
section can be used to obtain a rough estimate and the search range for
the unknown parameters, and
then the SNLS approach can be used to refine the estimates.

\section{Simulation studies}\label{sec4}

In this section we evaluate the performance of the MSSB and the SNLS
approaches based on
equations~(\ref{3Dmodel:Eq1})--(\ref{3Dmodel:Eq3}) by Monte Carlo
simulations. An ad hoc bandwidth
selection procedure is used for selecting a~proper bandwidth $h$ for
MSSB. See Liang and Wu (\citeyear{Liang2008})
for more detailed discussions.
The following parameter values were used to generate simulation data:
$T_U (0) = 600$, $T_I (0) = 30$, $V(0) = 10^5$, $\lambda= 36$, $\rho=
0.108$, $N=1000$, $\delta= 0.5$, $c=3$, and $\eta(t) = 9\times
10^{-5} \times\{1-0.9 \cos(\pi t/1000)\}$. Let $T=T_U +
T_I$ denote the total number of infected
and uninfected \mbox{CD4$+$ T} cells and $V$ denote the viral load, the
measurement models
(\ref{log10MeasurementModel})--(\ref{log10MeasurementModel4.3}) were
used to simulate the observation data with measurement noise, where
$\varepsilon_{1i}$ and $\varepsilon_{2j}$ were assumed to be
independent and
generated from normal distributions with mean zero and constant
variances $\sigma_{1}^2$ and~$\sigma_{2}^2$, respectively.
Equations~(\ref{3Dmodel:Eq1})--(\ref{3Dmodel:Eq3}) were numerically
solved within the time range $[0,20]$ (days) using the Runge--Kutta
method, and solutions were output for equally-spaced time intervals of
0.1 and 0.2 which correspond to the number of measurements 200 and 100,
respectively. The measurement errors $\sigma_{1}^2 = 20,30,40$ and
$\sigma_{2}^2 = 100,150,200$ were added to the numerical results of the
ODE model according to the measurement models (\ref
{log10MeasurementModel})--(\ref{log10MeasurementModel4.3}), respectively.

To evaluate the performance of the MSSB and SNLS approaches for smaller
samples sizes and larger variances that are similar to the actual
clinical data evaluated in the next section, we also performed the
simulations for the number of measurements $n=n_T=n_V=30$ and $50$, and
$\sigma_{1}^2 = 20^2,30^2,40^2$ and $\sigma_{2}^2 = 50^2,75^2,100^2$.
To evaluate the performance of the estimation methods, we use the
average relative estimation error (ARE) which is defined as
\[
\mathit{ARE}=\frac{1}{N} \sum_{j=1}^{N} \frac{|\vec{\theta}-\hat{\vec
{\theta
}}_j|}
{|\vec{\theta}|} \times100\%,
\]
where $\hat{\vec{\theta}}_j$ is the estimate of parameter $\vec
{\theta
}$ from the $j${th} simulation data set and $N$ is the total number of
simulation runs. We applied the MSSB and SNLS approaches to 500
simulated data sets for each simulation scenario. We used the AICc and
determined that the time-varying parameter $\eta(t)$ was approximated
by a spline of order 2 with 3 knots in our simulation studies.

The simulation results are reported in Table \ref{AREComparison}. From
these results, we can see that the SNLS approach significantly improves
the performance of the MSSB approach in all simulation cases as we
expected. When the number of measurements is large (e.g.,
$n=n_T=n_V=100$ or $200$) and variances are small, the SNLS approach
performs extremely well, but the AREs of the MSSB estimates of some
parameters such as $\rho$ and $\delta$ are very large. When the number
of measurements is small (similar to our real data analysis in the next
section, that is, $n=n_T=n_V=30$ or $50$) and the variances are large,
the performance of the SNLS estimates is still reasonably good, but the
MSSB approach performs poorly for all parameter estimates. Similarly,
the estimate of time-varying parameter $\eta(t)$ from the SNLS approach
outperforms that of the MSSB approach. Thus, the simulation results and
our experience suggest that it is a good strategy to use the MSSB
approach to obtain rough search ranges for unknown parameters, and then
use the SNLS approach to refine the estimates.

\begin{table}
\tablewidth=\textwidth
\tabcolsep=0pt
\caption{Simulation study: a comparison of the MSSB and SNLS
approaches. The average relative error
(ARE) was calculated based on 500 simulation runs. The
time-varying
parameter~$\eta(t)$ is
approximated by a 3-knots spline of order 2} \label{AREComparison}
\begin{tabular*}{\textwidth}{@{\extracolsep{\fill
}}lt{4.0}t{5.0}d{3.2}d{3.2}d{3.2}d{3.2}d{2.2}d{2.2}d{2.2}d{1.2}d{2.2}d{1.2}@{}}
\hline
& & & \multicolumn{5}{c}{\textbf{ARE (\%): MSSB}}& \multicolumn
{5}{c}{\textbf{ARE (\%):
SNLS}}\\[-6pt]
& & & \multicolumn{5}{c}{\hrulefill}& \multicolumn{5}{c}{\hrulefill
}\\
$\bolds{\bolds{n}}$ & \multicolumn{1}{c}{$\bolds{\bolds{\sigma
_1^2}}$} & \multicolumn{1}{c}{$\bolds{\sigma_2^2}$} & \multicolumn
{1}{c}{$\bolds{\lambda}$}
& \multicolumn{1}{c}{$\bolds{\rho}$} & \multicolumn{1}{c}{$\bolds
{N}$} & \multicolumn{1}{c}{$\bolds{\delta}$} &
\multicolumn{1}{c}{$\bolds{c}$} & \multicolumn{1}{c}{$\bolds
{\lambda}$} & \multicolumn{1}{c}{$\bolds{\rho}$} & \multicolumn
{1}{c}{$\bolds{N}$}
& \multicolumn{1}{c}{$\bolds{\delta}$} & \multicolumn{1}{c}{$\bolds
{c}$} \\
\hline
\phantom{0}30& 400. & 2500. &183.70&277.74&104.55&110.60&92.37
&14.50&28.00&5.95&6.31&1.16 \\
& 900. & 5625. &345.03&537.24&105.48&130.67&93.26
&18.90&36.40&7.60&8.38&1.72 \\
& 1600. & 10\mbox{,}000. &411.94&668.88&112.47&139.69&94.96
&22.80&44.60&9.34&10.70&2.19\\[3pt]
\phantom{0}50& 400. & 2500. &77.41&99.53&81.93&308.40&65.46
&11.70&22.80&4.66&4.69&0.68\\
& 900. & 5625. &98.32&106.77&104.55&384.52&66.59
&15.20&29.40&6.10&6.41&0.95\\
& 1600. & 10\mbox{,}000. &267.91&259.13&112.00&458.14&67.86
&19.30&38.00&7.68&8.20&1.18\\[3pt]
100& 20. & 100 &24.37&40.89&15.70&65.18 &36.92
&2.50&5.89&1.54&1.16&0.12\\
& 30. & 150. &25.13&42.50&16.37&97.53&37.65 &2.84&6.44&1.68&1.14&0.13\\
& 40. & 200. &27.56&46.98&16.45&114.09&38.84
&2.96&7.19&1.72&1.18&0.15\\[3pt]
200& 20. & 100 &8.01&14.17&9.72&153.74&15.15
&1.59&4.55&1.52&0.61&0.09\\
& 30. & 150. &8.24&14.52&10.42&156.17&15.54 &2.32&4.97&1.06&0.84&0.10\\
& 40. & 200. &8.85&15.62&10.76 &156.83&16.25 &2.13&5.37&1.32&0.96&0.12\\
\hline
\end{tabular*}
\end{table}

%
\begin{table}[b]
\tablewidth=\textwidth
\caption{Model selection for Patient I: the time-varying parameter
$\eta(t)$ is approximated\break using splines of order 3 or 4 with
the number of knots from 3 to 10} \label{PatientOneModelSelect}
\begin{tabular*}{279pt}{@{\extracolsep{\fill}}lccd{4.1}d{4.1}d{4.1}@{}}
\hline
\textbf{Model} &\textbf{Spline order} & \textbf{Number of knots} &
\multicolumn{1}{c}{\textbf{AIC}} & \multicolumn{1}{c}{\textbf{BIC}}
& \multicolumn{1}{c@{}}{\textbf{AICc}} \\
\hline
\phantom{0}1 & 3 & \phantom{0}3 & -165.7 & -151.5 & -161.7 \\
\phantom{0}2 & & \phantom{0}4 & -163.8 & -147.7 & -158.8 \\
\phantom{0}3 & & \phantom{0}5 & -174.4 & -156.5 & \multicolumn
{1}{c}{\phantom{1}$\textbf{$-$168.4}$} \\
\phantom{0}4 & & \phantom{0}6 & -173.6 & -154.0 & -165.6 \\
\phantom{0}5 & & \phantom{0}7 & -172.9 & -151.4 & -162.9 \\
\phantom{0}6 & & \phantom{0}8 & -177.0 & -153.8 & -165.0 \\
\phantom{0}7 & & \phantom{0}9 & -177.1 & -152.2 & -163.1 \\
\phantom{0}8 & & 10 & -173.9 & -147.1 & -156.9 \\[3pt]
\phantom{0}9 & 4 & \phantom{0}3 & \multicolumn{1}{c}{--} &
\multicolumn{1}{c}{--} & \multicolumn{1}{c}{--} \\
10 & & \phantom{0}4 & -170.5 & -154.4 & -165.5 \\
11 & & \phantom{0}5 & -173.7 & -155.8 & -167.7 \\
12 & & \phantom{0}6 & -173.4 & -153.7 & -165.4 \\
13 & & \phantom{0}7 & -171.5 & -150.1 & -161.5 \\
14 & & \phantom{0}8 & -172.3 & -149.1 & -160.3 \\
15 & & \phantom{0}9 & -175.5 & -150.6 & -161.5 \\
16 & & 10 & -174.0 & -147.3 & -157.0 \\
\hline
\end{tabular*}
\end{table}
%


%
\begin{table}
\tablewidth=\textwidth
\caption{Model selection for Patient II: the time-varying parameter
$\eta(t)$ is approximated\break using splines of order 3 or 4 with
the number of knots from 3 to 10} \label{PatientTwoModelSelect}
\begin{tabular*}{281pt}{@{\extracolsep{\fill}}lccd{4.1}d{4.1}d{4.1}@{}}
\hline
\textbf{Model} &\textbf{Spline Order} & \textbf{Number of knots} &
\multicolumn{1}{c}{\textbf{AIC}} & \multicolumn{1}{c}{\textbf{BIC}}
& \multicolumn{1}{c@{}}{\textbf{AICc}} \\
\hline
\phantom{0}1 & 3& \phantom{0}3 & -246.5 & -229.1 & -244.5 \\
\phantom{0}2 & & \phantom{0}4 & -244.5 & -225.0 & -241.5 \\
\phantom{0}3 & & \phantom{0}5 & -250.7 & -229.0 & \multicolumn
{1}{c}{\phantom{0}$\textbf{$-$246.7}$} \\
\phantom{0}4 & & \phantom{0}6 & -245.0 & -221.1 & -241.0 \\
\phantom{0}5 & & \phantom{0}7 & -252.4 & -226.3 & -246.4 \\
\phantom{0}6 & & \phantom{0}8 & -247.3 & -219.1 & -240.3 \\
\phantom{0}7 & & \phantom{0}9 & -249.5 & -219.0 & -241.5 \\
\phantom{0}8 & & 10 & -247.9 & -215.3 & -238.9 \\[3pt]
\phantom{0}9 &4 & \phantom{0}3 & \multicolumn{1}{c}{--} &
\multicolumn{1}{c}{--} & \multicolumn{1}{c}{--} \\
10 & & \phantom{0}4 & -245.7 & -226.2 & -242.7 \\
11 & & \phantom{0}5 & -248.5 & -226.7 & -244.5 \\
12 & & \phantom{0}6 & -243.9 & -220.0 & -239.9 \\
13 & & \phantom{0}7 & -249.8 & -223.7 & -243.8 \\
14 & & \phantom{0}8 & -245.7 & -217.4 & -238.7 \\
15 & & \phantom{0}9 & -247.1 & -216.6 & -239.1 \\
16 & & 10 & -244.0 & -211.4 & -235.0 \\
\hline
\end{tabular*}
\end{table}

\begin{figure}

\includegraphics{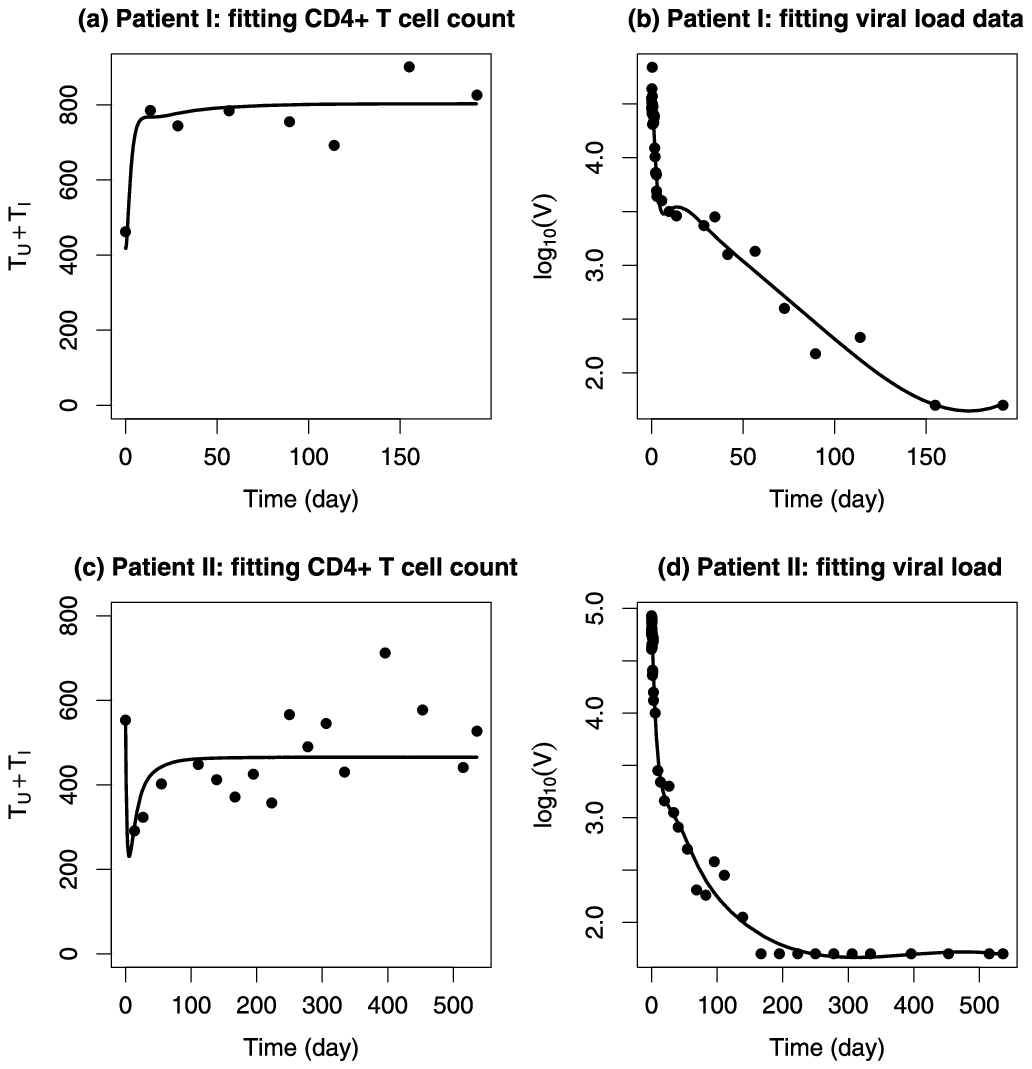}

\caption{The SNLS model fitting for two AIDS patients.}
\label{PatientOneTwoCurves}
\end{figure}

\section{Viral dynamic parameter estimation for AIDS patients}\label
{SecClinicalData}

We applied the proposed MSSB and SNLS approaches to estimate viral
dynamic parameters from data for individual AIDS patients based on
model equations~(\ref{3Dmodel:Eq1})--(\ref{3Dmodel:Eq3}). Two \mbox{HIV-1}
infected patients were treated with a four-drug antiretroviral regimen
in combination with an immune-based therapy. Frequent viral load
measurements were scheduled at baseline and after initiating the
combination treatment: 13 measurements during the first day, 14
measurements from day 2 to week 2, and then one measurement at each of
the following weeks, 4, 8, 12, 14, 20, 24, 28, 32, 36, 40, 44, 48, 52,
56, 64, 74 and 76, respectively. Total \mbox{CD4$+$ T} cell counts were
monitored at weeks 2, 4 and monthly thereafter.

As suggested in Section~\ref{multi-stage}, first we applied the
nonlinear regression model in Perelson et al. (\citeyear{Perelson1996}) to fit the
viral load data for the first week to obtain the estimates of
$\delta$ and $c$. The estimation results for the two patients are as follows:
Patient~I, $\delta= 1.09$ and $c = 2.46$; and Patient~II, $\delta=
0.43$ and $c=3.78$. Then we applied the proposed MSSB and SNLS
approaches to estimate all other viral dynamic parameters including
the time-varying parameter $\eta(t)$ in
models~(\ref{3Dmodel:Eq1})--(\ref{3Dmodel:Eq3}). The $\log_{10}$ transformation of the data was
used in order to stabilize the variance and computational
algorithms.

A rough estimate and a reasonable search range for each of the unknown
parameters and the initial values of the state variables were obtained
using the MSSB method. Then we applied the SNLS approach to refine the
estimates. For the SNLS approach, we selected the smoothing parameters
using the AIC, BIC and AICc \mbox{criteria}. We considered all the
combinations of two spline orders (3 and 4) and~8 different numbers of
control points (from 3 to 10), and the results are reported in
\mbox{Tables}~\ref{PatientOneModelSelect} and \ref{PatientTwoModelSelect}
for Patients I and II, respectively. Note that as a practical
guideline, if the number of unknown parameters exceeds $n/40$ (where
$n$ is the number of measurements), the AICc instead of AIC should be
used. For our clinical data, $n$ is about 40, and the number of unknown
parameters varies between~8 and 15 for different models (the unknown
initial conditions were also considered as unknown parameters), which
is much larger than $n/40 = 40/40 = 1$. So the AICc is more appropriate
for our case. In fact, the AICc converges to AIC as the number of
measurements gets larger, thus, the AICc is often suggested to be
employed regardless of the number of measurements [Burnham and Anderson
(\citeyear{Burnham2004})]. Therefore, our model selection is mainly based on AICc,
although the AIC and BIC scores are also reported in Tables~\ref
{PatientOneModelSelect} and \ref{PatientTwoModelSelect} for
comparisons. From Table~\ref{PatientOneModelSelect}, we can see that
both AICc and BIC selected the best spline model for $\eta(t)$ as order~3 and 5 control points for Patient I, although the AIC selected a
different model. From Table~\ref{PatientTwoModelSelect}, we can see
that the AICc selected the same
spline model (order~3 with~5 control points) for Patient II and both
AIC and BIC ranked this model as the second best model. Based on above
discussions, we are confident that the spline model with order 3 and~5
control points is the best approximation to the time-varying parameter
$\eta(t)$. Thus, we will mainly report and discuss the results from
this model.\looseness=1

\begin{table}
\tabcolsep=0pt
\tablewidth=\textwidth
\caption{The estimated values and associated 95\% confidence
intervals (CI) of viral dynamic parameters for two AIDS patients.
Perelson\textup{'}s model [Perelson et al. (\protect\citeyear{Perelson1996})] was first
fitted to the
viral load data within the first week to obtain estimates of
$\delta$ and $c$. For Patient I, $\delta= 1.09$ and $c = 2.46$; for
Patient~II, $\delta= 0.43$ and $c=3.78$. The time varying
parameter $\eta(t)$ of both Patients I and II was approximated using
a spline of order 3 with 5 control points}\label{ParaEstAllPatients}
\begin{tabular*}{\textwidth}{@{\extracolsep{\fill}}ld{4.2}d{3.2}d{4.2}d{4.2}@{}}
\hline
 & \multicolumn{2}{c}{\textbf{Patient
I}}&\multicolumn{2}{c@{}}{\textbf{Patient
II}}\\[-6pt]
& \multicolumn{2}{c}{\hrulefill}&\multicolumn{2}{c@{}}{\hrulefill}\\
& \multicolumn{1}{c}{\textbf{MSSB}} & \multicolumn{1}{c}{\textbf{SNLS }} & \multicolumn{1}{c}{\textbf{MSSB}}
& \multicolumn{1}{c@{}}{\textbf{SNLS}} \\
\textbf{Parameter}& \multicolumn{1}{c}{\textbf{(CI)}} &\multicolumn{1}{c}{\textbf{(CI)}}& \multicolumn{1}{c}{\textbf{(CI)}}&
\multicolumn{1}{c}{\textbf{(CI)}}\\
\hline
$\lambda$ (cell per day) &254.49 & 397.09 &18.38 &45.45\\
& \multicolumn{1}{c}{(181.11, 327.87)}& \multicolumn{1}{c}{(216.43, 594.30)}&\multicolumn{1}{c}{(3.50, 33.27)}&\multicolumn{1}{c}{(29.78,
81.48)}\\[3pt]
$\rho$ (per day) &0.34&0.49
&0.04&0.10 \\
& \multicolumn{1}{c}{(0.24, 0.44)}& \multicolumn{1}{c}{(0.26, 0.75)}&\multicolumn{1}{c}{(0.00, 0.08)}&\multicolumn{1}{c}{(0.06,
0.18)}\\[3pt]
$N$ (virion per cell) &1178.23 &264.74 &2769.32 &1114.37 \\
&\multicolumn{1}{c}{(558.20, 1798.26)}&\multicolumn{1}{c}{(203.40, 350.00)}&\multicolumn{1}{c}{(2484.54, 3074.10)}&\multicolumn{1}{c}{(856.62,
1428.93)}\\
\hline
\end{tabular*}
\end{table}

Figure~\ref{PatientOneTwoCurves} shows the data and model fitting
results from the best SNLS approach for the two patients. The fitted
curves are reasonably well. Table~\ref{ParaEstAllPatients} reports the
estimation results from both MSSB and SNLS methods for the two
patients. The MSSB approach provided good search ranges of unknown
parameters for the SNLS approach. From the SNLS estimates, the
proliferation rates of uninfected \mbox{CD4$+$ T} cells are $\lambda=397$ and
$45$ cells per day, respectively for Patients I and II; the death rates
are $\rho=0.49$ and $0.10$ per day which
correspond to the half-life of 1.4 and 6.9 days for the two patients,
respectively; the numbers of virions produced by each of the infected
cells are 265 and 1114 per cell for Patients I and II, respectively.

Note that the nonlinear regression estimates of the death rate of
infected cells,~$\delta$, for the two patients are 1.09 and 0.43 per
day with the corresponding half-life of 0.64 and 1.61 days,
respectively, which indicates a higher death rate of infected cells
compared to the uninfected cells for both patients. The estimates of
viral clearance rate ($c$) were 2.46 and 3.78 per day with the
corresponding half-life of 0.28 and 0.18 days for the two patients,
respectively. These results are
similar to those from the previous studies [Perelson et al. (\citeyear{Perelson1996},
\citeyear{Perelson1997})] and biologically reasonable. However, to our knowledge, this is
the first time the estimates of the proliferation rate and death rate
of uninfected \mbox{CD4$+$ T} cells ($\lambda$ and $\rho$) and the number of
virions produced by an infected cell ($N$) have been obtained directly
from the clinical data of AIDS patients. From the estimation results in
Table~\ref{ParaEstAllPatients}, we can see that the death rate of
infected cells is 2 to 4 fold higher than that of uninfected cells for
Patients I and II, respectively. We can also see that the difference in
the estimated dynamic parameters between the two patients is large.
This is consistent with the argument that the between-subject variation
of viral dynamics in AIDS patients is large [Wu and Ding (\citeyear{WuDing1999}), Wu et
al. (\citeyear{Wuetal1999})]. Thus, the personalized treatment is necessary for AIDS
patients.

\begin{figure}[b]

\includegraphics{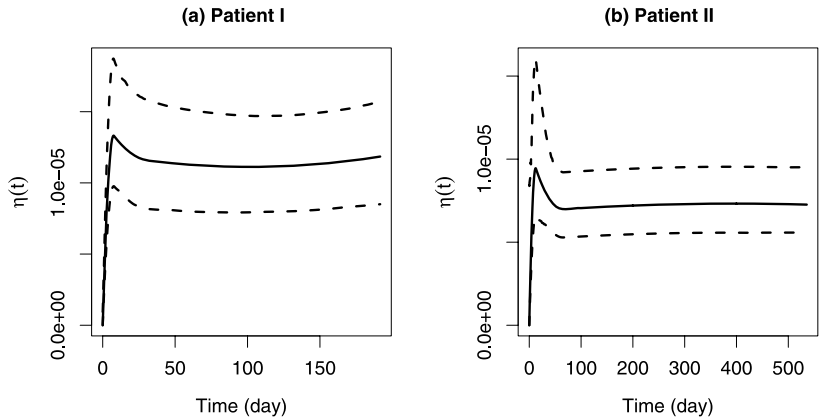}

\caption{The
estimated infection rate $\eta(t)$ (solid line) and its bootstrap 95\%
confidence interval (dashed
line).} \label{TwoPatientEta}
\end{figure}

The estimated time-varying parameter, the infection rate $\eta(t)$ for
Patients I and~II are plotted in Figure~\ref{TwoPatientEta}. From the
estimates, we can see that the infection rates for the two patients
have similar patterns which are primarily due to similar patterns of
viral load and \mbox{CD4$+$ T} cell counts for the two patients. It seems that
there was an initial fluctuation in the infection rate at the beginning
of treatment, and then the infection rate was stabilized after one or two
weeks. We can also see that the estimated infection rate is not zero,
which may suggest the imperfectness of the long-term treatment
and the development of drug-resistant mutants. This is inconsistent
with the perfect treatment assumption in Perelson et al. (\citeyear{Perelson1996,Perelson1997}),
although it may be valid in a short period of time after initiating the
treatment.

\section{Discussion and conclusion}\label{sec6}

We have proposed two approaches to identify all dynamic parameters
including both constant and time-varying parameters in an HIV viral
dynamic model which is characterized by a set of nonlinear differential
equations. This is a very challenging problem in the history of HIV
dynamic studies. The proposed multistage smoothing-based (MSSB)
approach is straightforward and easy to implement since it does not
require numerically solving the ODEs and the initial values of the
state variables are not needed. However, one limitation of this
approach is that it requires the estimates of derivatives of the state
variables, which are usually poor when the data are sparse. This may
result in poor estimates of unknown dynamic parameters. The second
method that we proposed is the spline-enhanced nonlinear least squares
(SNLS) approach. This method is more accurate in estimating the unknown
parameters, but it requires numerical evaluations of ODEs repeatedly in
the optimization procedure and the convergence of the computational
algorithm is problematic when the parameter space is high-dimensional.
We have proposed a hybrid optimization technique to deal with the
nonconvergence and the local solution problems, but the computational
cost is high and a good search range for each of the unknown parameters
is required. In practical implementation, we propose to combine the two
approaches, that is, first using the MSSB approach to obtain rough
estimates and possible ranges for the unknown parameters and then using
the SNLS approach to refine the estimates. We have applied this
strategy to estimate all dynamic parameters from data for two AIDS patients.

It is very important to estimate all dynamic parameters in HIV dynamic
models directly from clinical data when prior knowledge is not
available or fixing some parameters will significantly affect the
estimates of other parameters in a dynamic model. To our knowledge,
this is the first attempt to propose a procedure to estimate both
constant and time-varying parameters in the proposed HIV dynamic model
directly from clinical data. Although Huang, Liu and Wu (\citeyear{Huangetal2006}) have
attempted to do so using a Bayesian approach, strong priors were used
for most of the dynamic parameters and a parametric form for the
time-varying parameter was employed. Notice that the time-varying
parameter $\eta(t)$ is a function of antiviral treatment effect in the
HIV dynamic model [Huang, Liu and Wu (\citeyear{Huangetal2006})]. It is very important to
estimate this time-varying parameter for each AIDS patient individually
in order to better design the treatment strategy and personalize the
treatment for each individual. In addition, to better assess the
treatment efficacy, the time-varying infection rate $\eta(t)$ can be
further modeled as $\eta(t)=\eta_{\mathit{pre}} \times(1-\eta_{\mathit{treat}} (t))$,
where $\eta_{\mathit{pre}}$ denotes the infection rate at the pre-treatment
equilibrium, and $\eta_{\mathit{treat}} (t)$ the time-varying infection rate
after treatment. By comparing $\eta_{\mathit{pre}}$ with $\eta(t)$, it will be
easy to quantify the treatment effects on infection rate. However,
since $\eta_{\mathit{pre}}$ is strongly correlated with $\eta_{\mathit{treat}} (t)$,
$\eta
_{\mathit{pre}}$ has to be fixed to estimate $\eta_{\mathit{treat}} (t)$. Unfortunately,
the pre-treatment equilibrium data were not collected in this study
such that we could not determine the value of $\eta_{\mathit{pre}}$ and
therefore to do the comparison. Finally, the treatment could also
affect parameter $N$; however, limited by the clinical measurements, we
simplified our model without considering parameter $N$ as time-varying.

We used both viral load and \mbox{CD4$+$ T} cell count data in order to identify
all dynamic parameters in the HIV dynamic models~(\ref
{3Dmodel:Eq1})--(\ref{3Dmodel:Eq3}). Usually AIDS investigators believe
that \mbox{CD4$+$ T} cell count data are very noisy with a large variation
due to both measurement errors and natural variations of \mbox{CD4$+$ T} cell
counts. Thus, it is important to improve the data quality in order to
get more reliable estimates of dynamic parameters, although it is
beyond statisticians' control. Our study also suggests that the
frequent measurements, in particular, the viral load measurements
during the early stage after initiating an antiviral therapy, are
important to identify some constant dynamic parameters such as $c$ and
$\delta$, and long-term monitoring of viral load and \mbox{CD4$+$ T} cell counts
is necessary to estimate other parameters, in particular, the
time-varying parameter.

\begin{appendix}

\section*{Appendix: Differential evolution and scatter search} \label{app}

The differential evolution (DE) algorithm [Storn and Price (\citeyear{Storn1997})] is a
typical evolutionary algorithm that searches the optimum by
inheritance, mutation, selection and crossover of parent populations
(e.g., vectors of possible parameter values in the parameter space).
The initial population is randomly generated from a uniform
distribution within the search range to cover the whole region, denoted
by $x_{i,G}$ $(i=1,2,\ldots,N_P)$, where $N_P$ is the number of members in
this generation. The subsequent population inherits and mutates by
randomly mixing the previous
generation with certain weights. Storn and Price (\citeyear{Storn1997}) recommended a
mutation method
\begin{equation} \label{DEmutation}
v_{i,G+1} = x_{r_1,G}+F(x_{r_2,G}-x_{r_3,G}),
\end{equation}
where $v_{i,G+1}$ is the $i${th} member of generation $(G+1)$,
$x_{r_1,G}$, $x_{r_2,G}$ and $x_{r_3,G}$ the members of the previous
generation $G$, and $F>0$ the amplification factor. Integers $r_1$,
$r_2$ and $r_3$ are randomly chosen from $\{ 1,2,\ldots,N_P \}$
which are mutually different from each other and different from $i$.
Also, to increase diversity, the mutated member
$v_{i,G+1}$ exchanges its components with $x_{i,G}$ with a given
probability, the crossover ratio. For this purpose, a number within
$[0,1]$ is generated for each component of $v_{i,G+1}$ by a uniform
random number generator. If this number is greater than the crossover
ratio, then the component of $x_{i,G}$ is kept. Finally, the best
member in the mutated population is selected by comparing the values of
the objective function. The convergence rate of the differential
evolution method depends on specific mutation strategies used, that is,
the amplification factor and the
crossover ratio. This method has been successfully employed in previous studies
[Miao et al. (\citeyear{Miao2008,Miao2009})].
For more details about the DE algorithm, the reader is referred to Storn
and Price (\citeyear{Storn1997}).

The scatter search method was first proposed by Glover (\citeyear{Glover1977}) and
extended later by Laguna and Marti (\citeyear{Laguna2003,Laguna2005}). The scatter search
method is not a genetic algorithm; instead, it locates the optimum
by tracking the search history and balancing visiting frequency within
different segments of the search region using sophisticated
strategies. Here we give a brief introduction to this method. For
more details, the reader is referred to Laguna and Marti (\citeyear{Laguna2003}) and
Rodriguez-Fernandez, Egea and Banga (\citeyear{Rodriguez2006}).

Let $l_{i}$ and $u_{i}$ denote the lower and upper boundary of the
$i${th} component of the parameter vector, respectively. Then the
region between $l_{i}$ and $u_{i}$ can be divided into $m$ segments
(e.g., $m=4$), which are usually of equal length. Let $s_{ij}\
(j=1,2,\ldots,m)$ denote the $j${th} segment of the search region
$[l_{i},u_{i}]$. If a parameter value candidate of
$\theta_i$ falls in $s_{ij}$, we call it a visit to $s_{ij}$. Let
$f_{ij}$ denote the total number of visits to $s_{ij}$, then a
visiting history matrix $\mathbf{F}$ can be formed:
\begin{equation} \label{chap4SSvisitmatrix}
\mathbf{F} = \left[\matrix{
f_{11} & f_{12} & \cdots& f_{1m}\vspace*{2pt} \cr
f_{21} & f_{22} & \cdots& f_{2m} \cr
\vdots& \vdots& \vdots& \vdots\vspace*{2pt}\cr
f_{q1} & f_{q2} & \cdots& f_{qm}}
\right].
\end{equation}
Then we can calculate the probability of historic visits to $s_{ij}$ as
\begin{equation}
\label{chap4SSVisitProb}
p_{ij} = \frac{{1}/{f_{ij}}}{\sum_{k=1}^m
{1}/{f_{ik}}},\qquad  i=1,2,\ldots,q;\mbox{ } j=1,2,\ldots,m.
\end{equation}
Since the scatter search method is population-based, the first
population needs to be generated to start the algorithm. Let $N_G$
denote the total number of parameter vectors in the first
population, then $m$ parameter vectors are generated first such that
the $i${th} component of the $j${th} vector falls into $s_{ij}$.
In this way, all elements of matrix $\mathbf{F}$ become one. To
generate the rest $(N_G-m)$ parameter vectors in the first
population, a random vector $z = (z_1, z_2, \ldots, z_q)^{\T}$ is
generated for each parameter vector with $z_i$ $(i=1,2,\ldots,q)$
following a uniform distribution on $[0,1]$. Then the~$i${th}
component of the parameter vector to be generated falls into
$s_{ik}$ for the smallest $k$ satisfying
\[
z_i \leq\sum_{j=1}^k p_{ij},\qquad  k=1,2,\ldots,m.
\]
Note that after each new parameter vector is generated, the matrix
$\mathbf{F}$ is also updated.

Once the first population is generated, the parameter vectors are
divided into two categories: \textit{elite vectors} and \textit{diverse
vectors}. Let $n_e$ denote the number of all elite vectors, which is
usually a small fixed number (e.g., 20). Then $n_e / 2$ elite vectors
are chosen based on the smallest objective function values; the rest
half of the elite vectors are chosen to be farthest from all the first
half elite vectors in the sense of Euclidean distances. Thus, both
fitness and
diversity are considered in the construction of the elite vectors. Then
the new parameter vectors can be generated by combination of the elite
vectors. Let $x^{(1)}$ and $x^{(2)}$ denote two different elite vectors
and $x^{(1)}$ has a smaller objective function value than $x^{(2)}$.
Then three types of combinations can be employed to generate new
parameter vectors:
\begin{itemize}
\item[$\bullet$]Type 1: $p_1 = x^{(1)} - d$,
\item[$\bullet$]Type 2: $p_2 = x^{(1)} + d$,
\item[$\bullet$]Type 3: $p_3 = x^{(2)} + d$,
\end{itemize}
where $d = r^{\T} \cdot(x^{(2)}-x^{(1)})$ with all the
components of $r$ generated from an uniform distribution on $[0,1]$. If
both $x^{(1)}$ and $x^{(2)}$ are in the first half elite vectors in
terms of fitness, one new vector of type 1, one of type 3 and two of
type 2 are generated; if only $x^{(1)}$ belongs to the first half elite
vectors in terms of fitness, one new vector of each type is generated;
if both $x^{(1)}$ and $x^{(2)}$ belong to the second half elite vectors
in terms of farthest distance, then one new vector of type 2 and one of
either type 1 or 3 are generated. The fitness of these new generated
vectors is then compared to that of the elite vectors. The new vectors
with smaller objective function values (that is, better fit) will
replace the elite vectors with larger objective function values. This
procedure is repeated until no new vectors can replace any elite
vectors. Now the algorithm can either stop or continue by regenerating
diverse vectors.
\end{appendix}
\section*{Acknowledgments}
The authors thank an Associate Editor and referees for their
constructive comments.

\printaddresses

\end{document}